\newcommand{\Tr}{\hbox{Tr}}
\documentclass[aps,pra,twocolumn,superscriptaddress,showpacs]{revtex4-1}
\usepackage{graphicx}
\usepackage{hyperref}
\usepackage{amsmath}
\usepackage{subfigure}
\usepackage{epsfig} 
\usepackage{epstopdf}
\usepackage{float} 
\usepackage{tikz}
\usepackage[percent]{overpic}
\usepackage{comment}
\usepackage{pstricks}
\usepackage{placeins}
\usepackage{multirow}
\usepackage{rotating}
\usepackage{hhline}

\begin{document}
\title{Effectiveness of classical spin simulations for describing NMR relaxation of quantum spins.}

\author{Tarek A. Elsayed}
\email{T.Elsayed@thphys.uni-heidelberg.de}
\address{Institute for Theoretical Physics, University of Heidelberg, Philosophenweg 19, 69120 Heidelberg, Germany}
\author{Boris V. Fine}
\email{B.Fine@thphys.uni-heidelberg.de}
\address{Institute for Theoretical Physics, University of Heidelberg, Philosophenweg 19, 69120 Heidelberg, Germany}
\address{Department of Physics, School of Science and Technology, Nazarbayev University
53 Kabanbai Batyr Ave., Astana 010000, Kazakhstan}
\address{Skolkovo Institute of Science and Technology, 100 Novaya Str., Skolkovo, Moscow Region 143025, Russia}

\date{\today}

\begin{abstract}
We investigate the limits of effectiveness of classical spin simulations for predicting free induction decays (FIDs) measured by solid-state nuclear magnetic resonance (NMR) on systems of quantum nuclear spins. The specific limits considered are associated with the range of interaction, the size of individual quantum spins and the long-time behavior of the FID signals. We compare FIDs measured or computed for lattices of quantum spins (mainly spins 1/2) with the FIDs computed for the corresponding lattices of classical spins. Several cases of excellent quantitative agreement between  quantum and classical FIDs are reported along with the cases of gradually decreasing quality of the agreement. We formulate semi-empirical criteria defining the situations, when classical simulations are expected to accurately reproduce quantum FIDs. Our findings indicate that classical simulations may be a quantitatively accurate tool of first principles calculations for a broad class of macroscopic systems, where individual quantum microscopic degrees of freedom are far from the classical limit.
\end{abstract}

\pacs{76.60.Pc, 76.60.-k, 31.15.xv, 03.65.Sq, 75.10.Pq}
\keywords{}
\maketitle

\section{Introduction}
\label{intro}

Predictive calculations of spin-spin relaxation in solid-state nuclear magnetic resonance (NMR) is a long-standing and still not fully solved problem\cite{VanVleck-48,Lowe-57,Abragam-61,Tjon-66,Gade-66,Borckmans-68,Parker-73,Jensen-73A,Engelsberg-75,Becker-76,Shakhmuratov-91,Lundin-92,Jensen-95,Fine-97,Zhang-07}. Similar problems also occur in the context of decoherence of solid-state qubits caused by nuclear spins\cite{Zhang-07}. Due to the smallness of nuclear gyromagnetic ratios, the limit of practical interest here is that of infinite temperature. At infinite temperature, the static equilibrium properties are trivial, but the dynamic ones are not. Direct numerical calculations of spin-spin relaxation are often not feasible, because the memory required grows exponentially with the number of spins in the system. In such a situation, simulations of classical spin lattices become an important computational resource.

It is common knowledge in the field of NMR that the dynamics of classical spins often well represents the behavior of systems of small quantum spins including even spins 1/2. This issue was investigated over the years by a number of authors\cite{Jensen-73,Jensen-76,Lundin-77,Tang-92,Elsayed-13-thesis}.  However, the limits of the accuracy of classical simulations for the description of quantum spin relaxation at high temperatures have not yet been established.  The goal of the present article is to investigate these limits as far as the the range of interaction, the size of individual quantum spins and the long-time behavior of spin-spin relaxation is concerned.

This work was in part motivated by the previous investigations of one of us\cite{Fine-04} that have shown that free induction decays (FIDs) in both classical and quantum spin systems exhibit generic exponential long-time decay on the time scale of microscopic spin-spin interaction. Theoretical analysis\cite{Fine-04}, numerical simulations\cite{Fabricius-97,Fine-03} and experiments\cite{Engelsberg-74,Morgan-08,Sorte-11,Meier-12} also indicate that, normally, the above long-time behavior becomes dominant after time of the order of characteristic spin-spin interaction time, i.e. rather fast. Therefore, we expected that, if the interaction constants for classical spin simulations are chosen such that the initial evolutions of the quantum and classical FIDs are matched, then the agreement between the two FIDs may last until the onset of the exponential long-time behavior, and, if so, the two FIDs will not diverge much afterwards. This expectation is largely confirmed by the results presented below even for the lattices of spins 1/2 with relatively few interacting neighbours.

The plan of the rest of the article is the following: Section~\ref{general} contains the formulation of the problem. In Section~\ref{experiment} , we compare classical spin calculations with the experimental NMR results for CaF$_2$. In Section~\ref{model} ,  we compare classical and quantum calculations for model spin systems.  Finally, Section~\ref{discussion}  contains a concluding discussion, which, in particular, includes semi-empirical criteria identifying quantum spin systems for which classical simulations are expected to produce quantitatively accurate FIDs.

\section{General formulation}
\label{general}

We consider translationally invariant spin lattices governed by the Hamiltonian:
\begin{equation}
\mathcal{H}=\sum_{m<n} J^x_{mn}  S_m^x S_n^x + J^y_{mn} S_m^y S_n^y + J^z_{mn}  S_m^z S_n^z ,
\label{H}
\end{equation}
where $S_m^{\alpha}$ represents either the quantum operator of the $\alpha^{\text{th}}$ ($x$, $y$ or $z$) projection of a quantum spin on $m^{\text{th}}$ lattice site or the corresponding projection of a vector of length 1 representing a classical spin,  $J^{\alpha}_{mn}$ are the coupling constants for the $\alpha^{\text{th}}$ projections of the $m$th and the $n$th spins. We use periodic boundary conditions. 

The quantity of interest for this study is an infinite temperature correlation function of the type that characterizes NMR free induction decay\cite{Lowe-57,Abragam-61}, namely:
\begin{equation}
C(t) \equiv \langle M_x(t) M_x (0) \rangle,
\label{Ct}
\end{equation}
where the $M_x \equiv \sum_n S_n^x$ is the total $x$-polarization of the system, and notation $\langle ... \rangle$ implies averaging over the infinite temperature equilibrium fluctuations. (When presenting the results, we always normalize $C(t)$ such that $C(0) = 1$.) For quantum systems, the above correlation function is calculated as 
\begin{equation}
C(t)=\Tr\{M_x(t)M_x\}=\Tr\{e^{i\mathcal{H}t}M_xe^{-i\mathcal{H}t}M_x\},
\label{Ct-quant}
\end{equation}
where the value of $\hbar$ is set to one. For classical systems, it is obtained as 
\begin{equation}
C(t)= \lim_{T\to \infty} \frac{1}{T} \int_0^{T} M_x(\tau) M_x(\tau+t) d\tau.
\label{Ct-class}
\end{equation}

In the quantum case, we compute the right-hand side (RHS) of Eq.(\ref{Ct-quant}) using a direct simulation of the time evolution of a randomly chosen initial wave function. The method is explained and rigorously justified in Ref.~\cite{Elsayed-13} on the basis of quantum typicality.  

In the classical case, the RHS of Eq.(\ref{Ct-class}) is obtained by the direct  simulations of classical spin dynamics governed by equations:
 \begin{equation}
  \dot{\bf{S}}_m=  {\bf{S}}_m \times {\bf{h}}_m ,
  \label{eom}
  \end{equation} 
  where 
  \begin{equation}
  {\bf{h}}_m \equiv \sum_n \left(
  \begin{array}{c} 
  J^x_{mn}  S_n^x \\
  J^y_{mn}  S_n^y  \\
  J^z_{mn}  S_n^z 
  \end{array}
    \right)
  \label{h}
  \end{equation}
 is the local field on the $m$th lattice site created by the neighbors.  The initial orientations of spins are chosen randomly. The simulations are based on a 4th-order Runge-Kutta algorithm.  Additional averaging is also performed over many  different realizations of random initial conditions.

The characteristic time scale of spin dynamics in both classical and quantum case can be characterized by the inverse root-mean-squared value of the local fields $h_m$:
\begin{equation}
\label{tau}
\tau = \left(
 \sum_n {J^x_{mn}}^{\!\!\!\! 2}   \left\langle {S_n^x}^2 \right\rangle +
  {J^y_{mn}}^{\!\!\!\! 2}  \left\langle {S_n^y}^2 \right\rangle  +
  {J^z_{mn}}^{\!\!\!\! 2}  \left\langle {S_n^z}^2 \right\rangle 
\right)^{- 1/2} .
\end{equation}
Whenever we compare quantum and classical lattices, the interaction constants of the classical Hamiltonian are equal to the interaction constants of the quantum Hamiltonian multiplied by factor $\sqrt{S (S+1)}$, where $S$ is the value of individual quantum spins involved.  Such a rescaling implies that the characteristic times $\tau$ are the same in both cases. It also guarantees that the second moments and hence the initial evolutions of the quantum and classical correlation functions are the same.

As far as the long-time behaviour of $C(t)$ is concerned,  the previous investigations \cite{Fine-04,Fabricius-97,Fine-03,Engelsberg-74,Morgan-08,Sorte-11,Meier-12} have shown that, in both classical and quantum systems, it has generic form
\begin{equation}
C(t) \cong e^{- \gamma t} \cos(\omega t + \phi),
\label{long-t}
\end{equation}
where $\gamma$ and $\omega$ are some constants, typically, of the order of $1/\tau$, and $\phi$ is an oscillation phase. Normally, this behaviour sets in after time of the order of $\tau$. In Refs.\cite{Fine-04}, the long-time behaviour (\ref{long-t}) was linked to the chaotic character of microscopic spin dynamics governed by Hamiltonian (\ref{H}).

\begin{figure*}[] \setlength{\unitlength}{0.1cm}
\begin{picture}(160 , 76 ) 
{\newcommand{\x} {9}
\put(-9, 39)  {\includegraphics[ width=5.5cm]{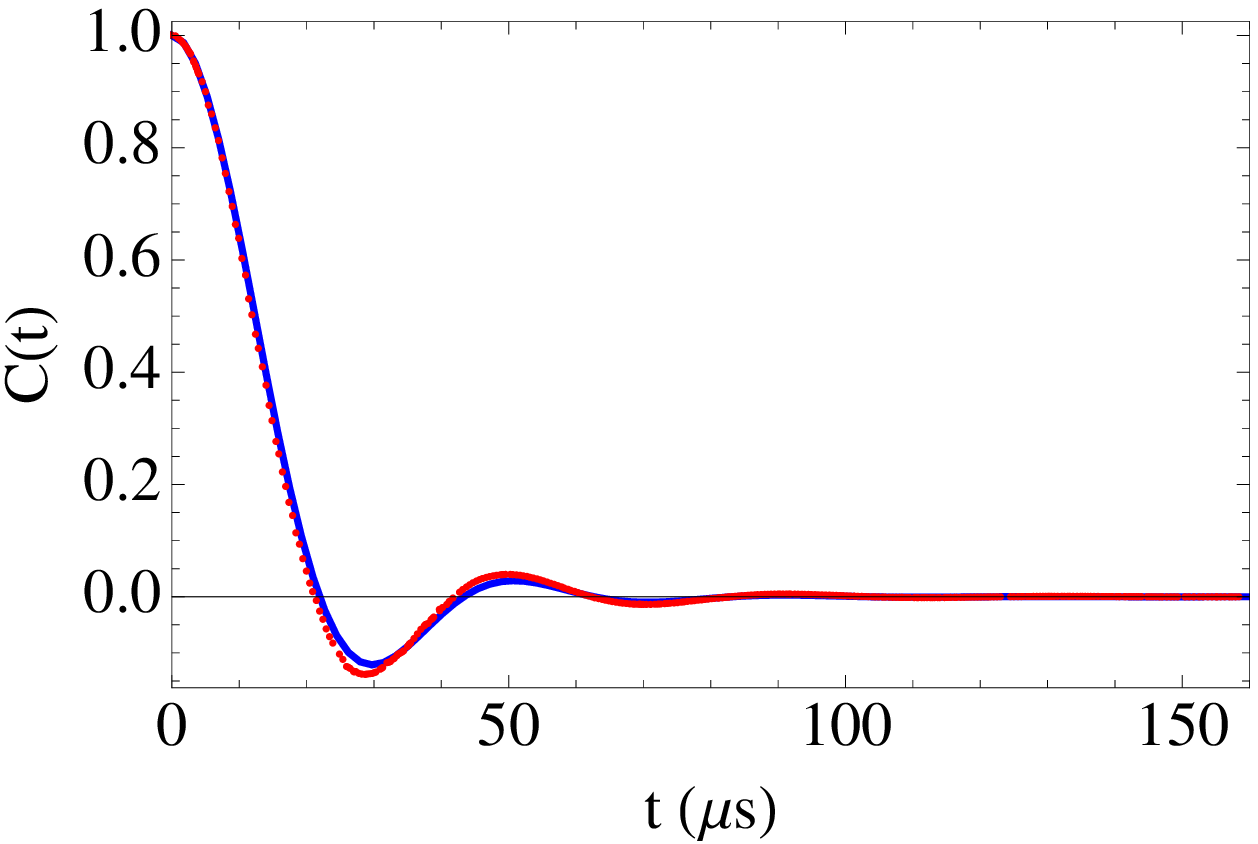}}
\put(51.5, 39)  {\includegraphics[ width=5.5cm] {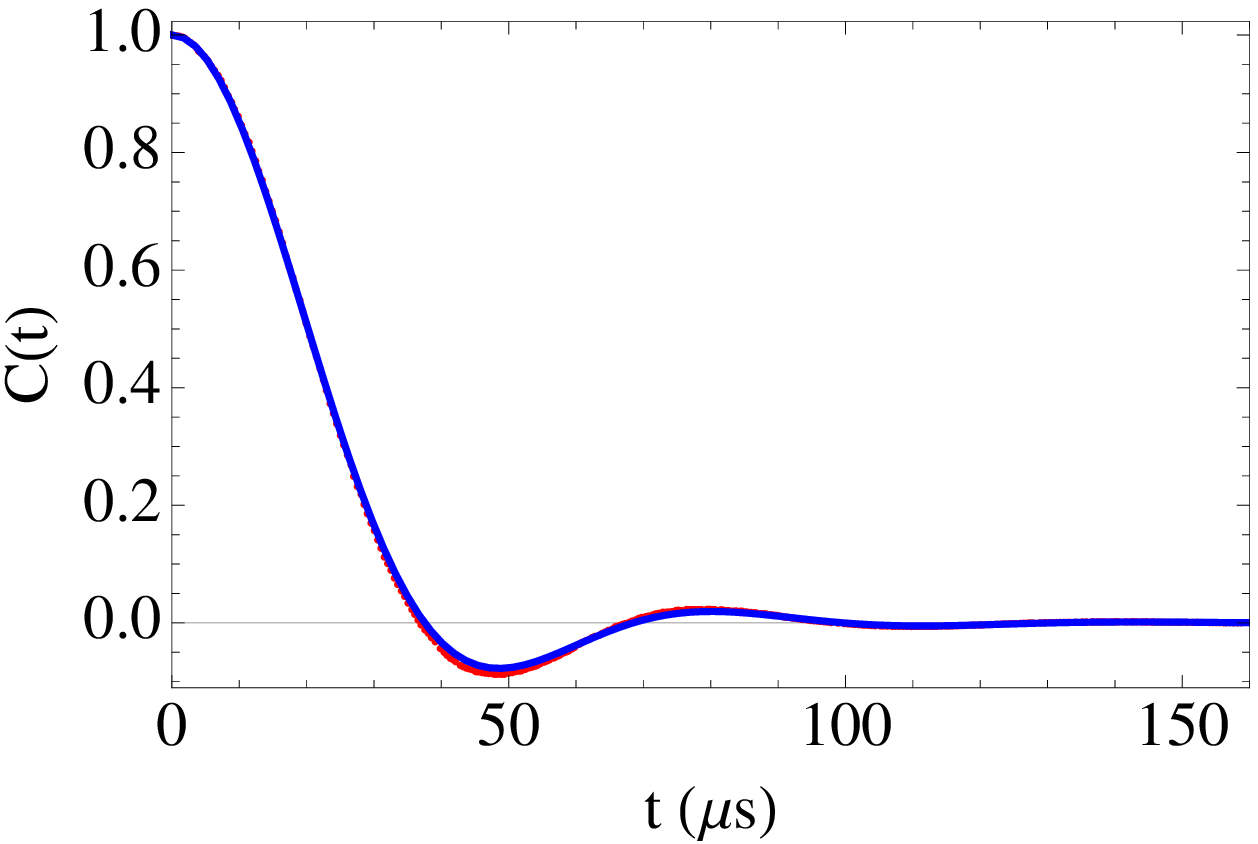}   }
\put(113, 39)  {\includegraphics[ width=5.5cm] {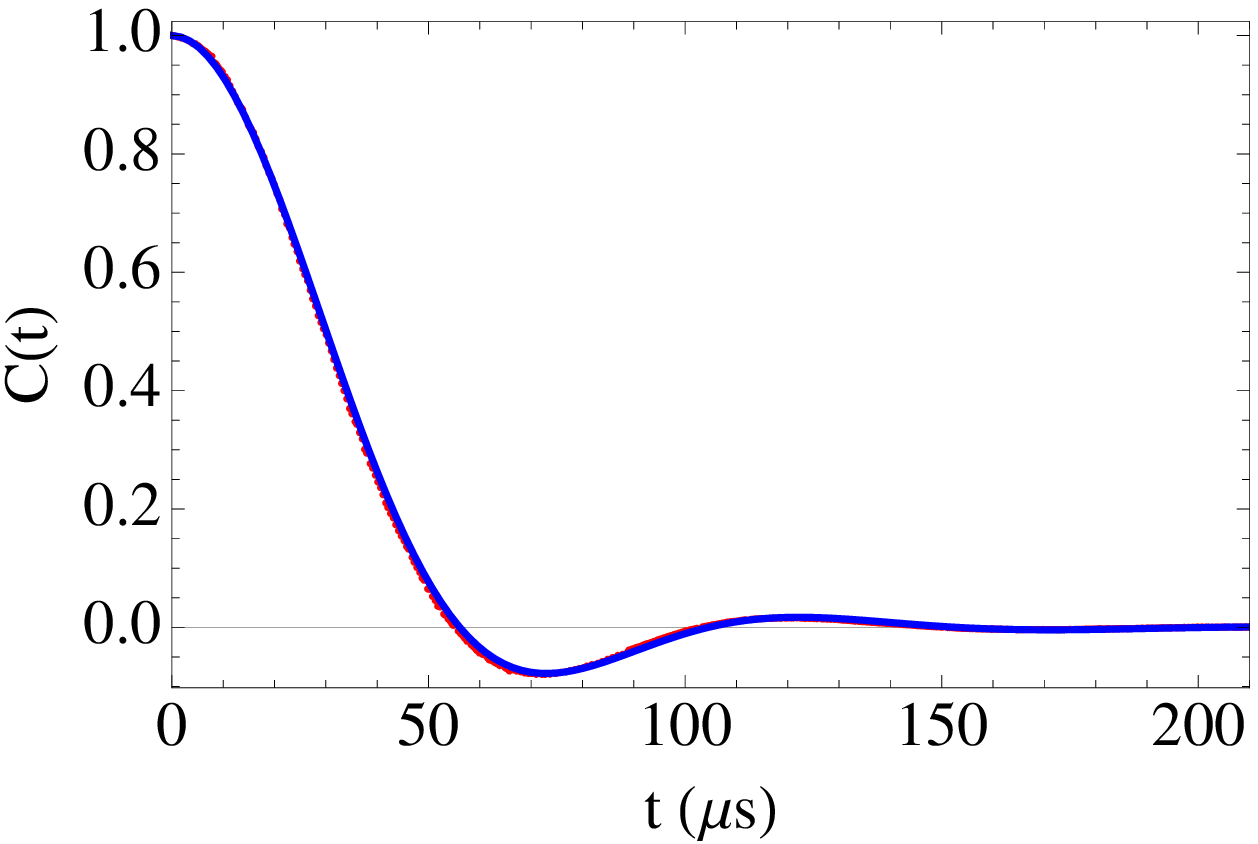} }

\put(-9, 0)  {\includegraphics[ width=5.5cm]{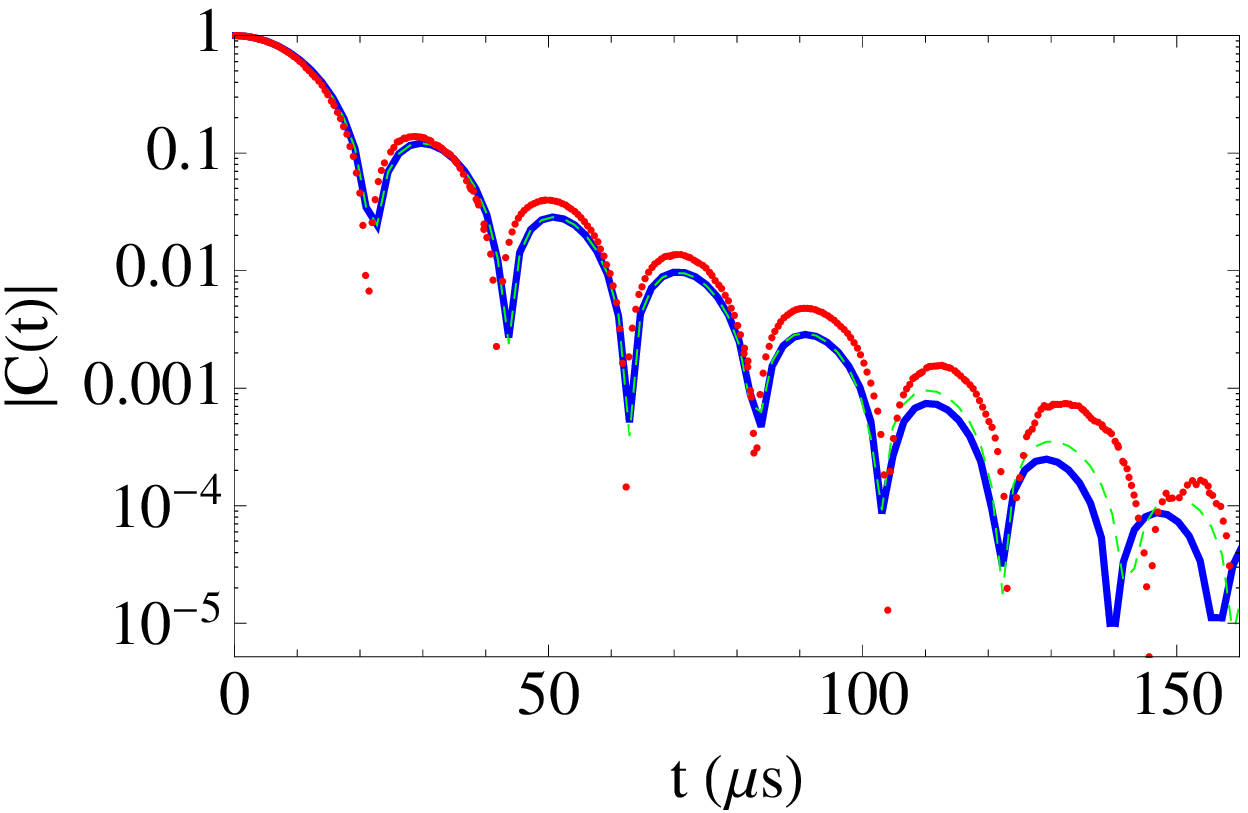}}
\put(51.5, 0)  {\includegraphics[ width=5.5cm] {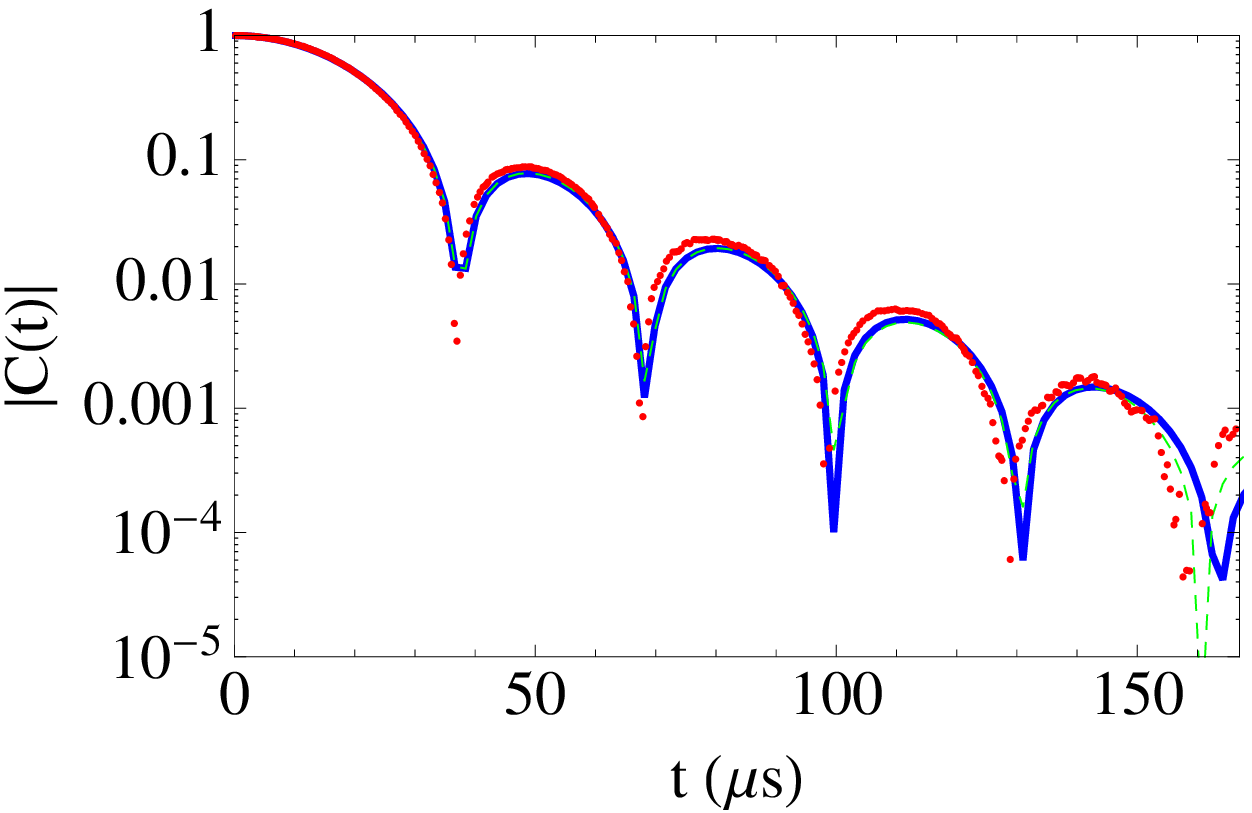}   }
\put(113, 0)  {\includegraphics[ width=5.5cm] {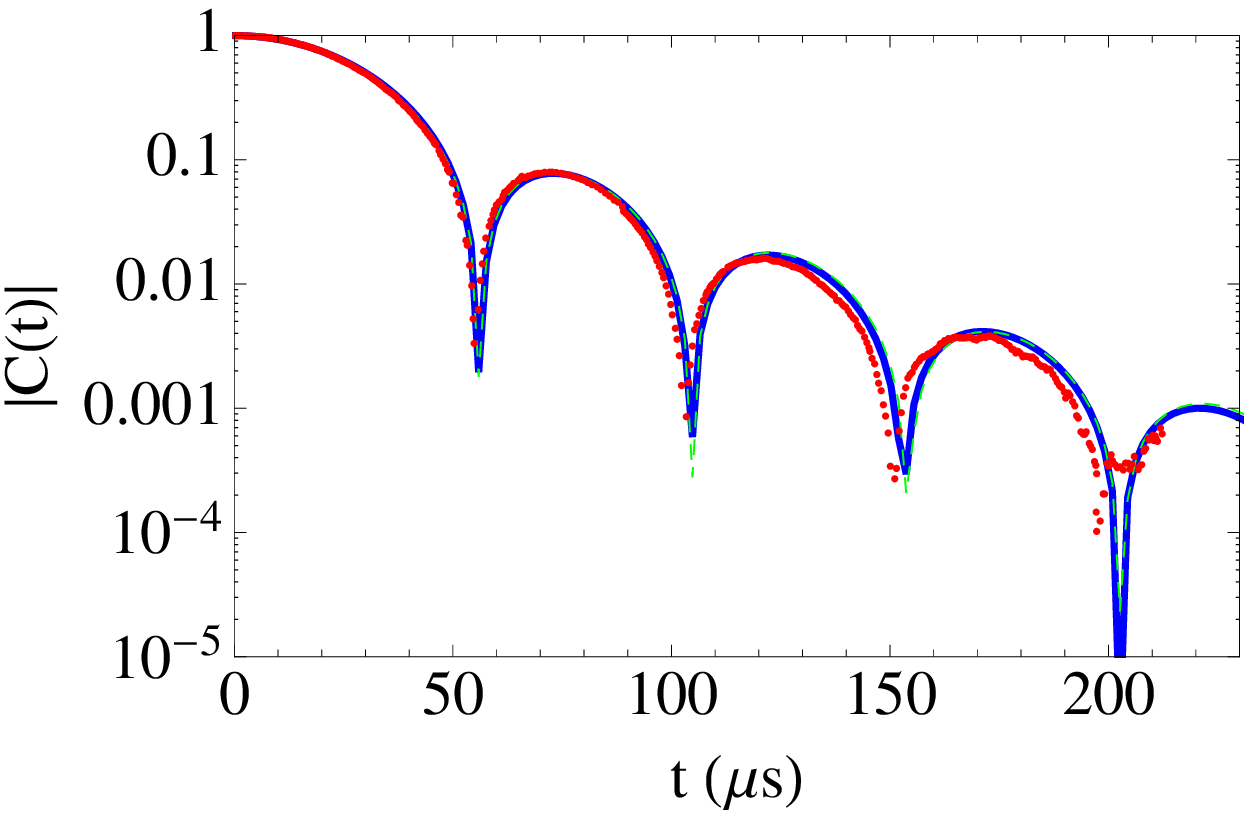} }

\put(-13,72) { {\bf(a) }}
\put(-13,32) { {\bf(b) }}

\put(49,72) { {\bf(c) }}
\put(49,32) { {\bf(d) }}

\put(110,72) { {\bf(e) }}
\put(110,32) { {\bf(f) }}

\put(35,29) { [100] }
\put(94,29) { [110] }
\put(154,29) { [111] }

\put(35,70) { [100] }
\put(94,70) { [110] }
\put(154,70) { [111] }
}
\end{picture} 

\caption{(Color online)  $^{19}$F FID for CaF$_2$. Red dots: experimental measurements extracted from Ref.~\cite{Engelsberg-74}. Blue solid lines: results of calculations for $11 \times 11 \times 11$ classical spin lattices. Green dashed lines: results for $9 \times 9 \times 9$ classical spin lattices (fully covered by blue lines in some plots). The external magnetic field is directed along the lattice directions [100], [111] and [110], as indicated on the plots.  Upper row: linear-scale plots. Lower row: semi-logarithmic plots. }
\label{class-FID}
\end{figure*}

\section{Free Induction decays in CaF$_2$}
\label{experiment}

CaF$_2$ is a benchmark material for testing theories of NMR spin-spin relaxation\cite{VanVleck-48,Lowe-57,Abragam-61}. Classical spin simulations of the $^{19}$F FIDs in CaF$_2$ have been done before\cite{Jensen-73,Tang-92,Zhang-07}. However, for various reasons such as different focus of investigation and/or drastic truncation of the range magnetic dipolar interaction,  the full quality of the agreement between the classical simulations and experimental results have not been fully exposed. Also the comparison of the long-time relaxation has not been made.

In CaF$_2$, fluorine nuclei form a simple cubic lattice with lattice period $a_0 = 2.72$~\AA. Fluorine has only one stable isotope $^{19}$F, which turns out to be magnetic. It has spin 1/2 with gyromagnetic ratio $g = 25 166.2$~rad~ s$^{-1}$~Oe$^{-1}$. Calcium nuclei are overwhelmingly nonmagnetic, so that their presence can be neglected.  The $^{19}$F  FID is measured as a relaxation of the total nuclear magnetization transverse to a strong magnetic field $\mathbf{B}_0$.  This relaxation is caused by the magnetic dipolar interaction between $^{19}$F spins. According to a linear response relation\cite{Lowe-57,Abragam-61,Elsayed-13}, the FID signal is proportional to the infinite temperature correlation function $C(t)$.

In the presence of a strong magnetic field, the full magnetic dipolar interaction should be truncated to keep only the terms that are preserved after averaging over the fast Larmor precession around the direction of the field\cite{VanVleck-48,Lowe-57,Abragam-61}. The resulting truncated Hamiltonian is conventionally presented in the Larmor rotating reference frame with the $z$-axis chosen parallel to the field. It has the general form (\ref{H}) with the interaction constants 
\begin{equation}
J^z_{mn}= -2 J^x_{mn} = -2 J^y_{mn} = \frac{g^2\hbar^2\left(1-3\cos^2 \theta_{mn}\right)}{|\mathbf{r}_{mn}|^3},
\label{J}
\end{equation}
where $\mathbf{r}_{mn}$ is the displacement vector between the $m$th and the $n$th lattice sites, and $\theta_{mn}$ is the angle between  $\mathbf{r}_{mn}$ and the external magnetic field $\mathbf{B}_0$.  Different orientations of $\mathbf{B}_0$ lead to different truncated Hamiltonians and hence different FIDs. We consider three orientations of $\mathbf{B}_0$ along [100], [110] and [111] crystal directions.

 We computed classical correlation functions $C(t)$ according to formula (\ref{Ct-class}) with $T=200 J^{-1}$, where $J=\gamma^2 \hbar^2 /a_0^3$ and with additional averaging over $3.2\times 10^{5}$ independent time evolutions.  The discretization time step was $\delta t=0.05J^{-1}$.  In these simulations, each spin interacted with each other with coupling constants(\ref{J}), where the vectors $\mathbf{r}_{mn}$ were determined as the shortest vectors connecting two lattice sites given the periodic boundary conditions.

Our results for $11 \times 11 \times 11$ lattice of classical spins are compared with the experimental results for $^{19}$F FID in CaF$_2$ in Fig.~\ref{class-FID}.  The above classical results were indistinguishable from those for $9 \times 9 \times 9$ lattice down to the values of $C(t) \sim 10^{-3}$. Therefore, we conclude that these results  are representative of the infinite-size lattices.

Both the simulated and the experimental FIDs exhibited the long-time behaviour of form (\ref{long-t}). The constants $\omega$ and $\gamma$ extracted by fitting Eq.(\ref{long-t}) to either the simulations or the experiment are compared with each other in Table~\ref{Table1}.

Overall, the agreement between the classical simulations and the experiment is excellent for the FIDs corresponding to the [110] and [111] directions of $\mathbf{B}_0$. The agreement for the [100] direction is also good but with a noticeable minor deviation as far as the linear-scale plot is concerned and then with larger deviation of the long-time tales.  

The latter discrepancy is consistent with larger differences of analytically computed moments for quantum and classical FIDs\cite{Jensen-73A,Jensen-76,Lundin-77,Tang-92}. It is also a likely consequence of the fact that the truncated Hamiltonian for $\mathbf{B}_0$ along the [100] direction  is such that each spin has only two strongest neighbors, while, for the four second-ranked neighbors, the coupling is two times smaller, i.e. the two strongest neighbors stand apart as far as the dynamic correlations are concerned. For comparison, the truncated Hamiltonian for $\mathbf{B}_0$ along the [111] direction also implies that each spin has two strongest neighbors, but then it has twelve second-ranked neighbors with coupling only eight percent smaller . 
 The above situation reflects the fact that magnetic dipole interaction is a transitional case between short-range and infinite-range interactions, which means that the short-range aspects of the interaction can play a noticeable role.

{\renewcommand{\arraystretch}{1.7}%
\begin{table}
\begin{center}
\begin{tabular}{|c|c|c|c|c|}

\cline{2-5} 
 
\multicolumn{1}{c|}{} &\multicolumn{2}{c|}{$\gamma$ (1/ms)} 	 & \multicolumn{2}{c|}{$\omega$ (rad/ms)}  \\

\cline{2-5} 

\multicolumn{1}{c|}{} &Experimental 	 & Numerical  &Experimental 	 & Numerical \\

\hline 
[100] & 50 & 60 & 151 & 154  \\
\hline
[110] & 42 &  44 & 103 & 101 \\
\hline
[111] & 29 & 31 & 66 & 65 \\
\hline
\end{tabular}
\end{center}
\caption{The values of $\gamma$ and $\omega$ obtained by fitting the functional dependence (\ref{long-t}) to the long-time behavior of the experimental and numerical FIDs  presented in Fig.~\ref{class-FID}.}
\label{Table1}
\end{table} 
}

\section{Model spin systems}
\label{model}

In this section, we compare correlation functions $C(t)$ for quantum and classical lattices with nearest-neighbor interactions. The lattices to be considered have different numbers of nearest neighbors and different quantum spin numbers $S$. Similar investigation for correlation functions related to spin diffusion were made in Ref.~\cite{Steinigeweg-12}.

\begin{figure} \setlength{\unitlength}{0.1cm}
\begin{picture}(88 , 106 )
{
\put(0, 54){ \epsfig{file=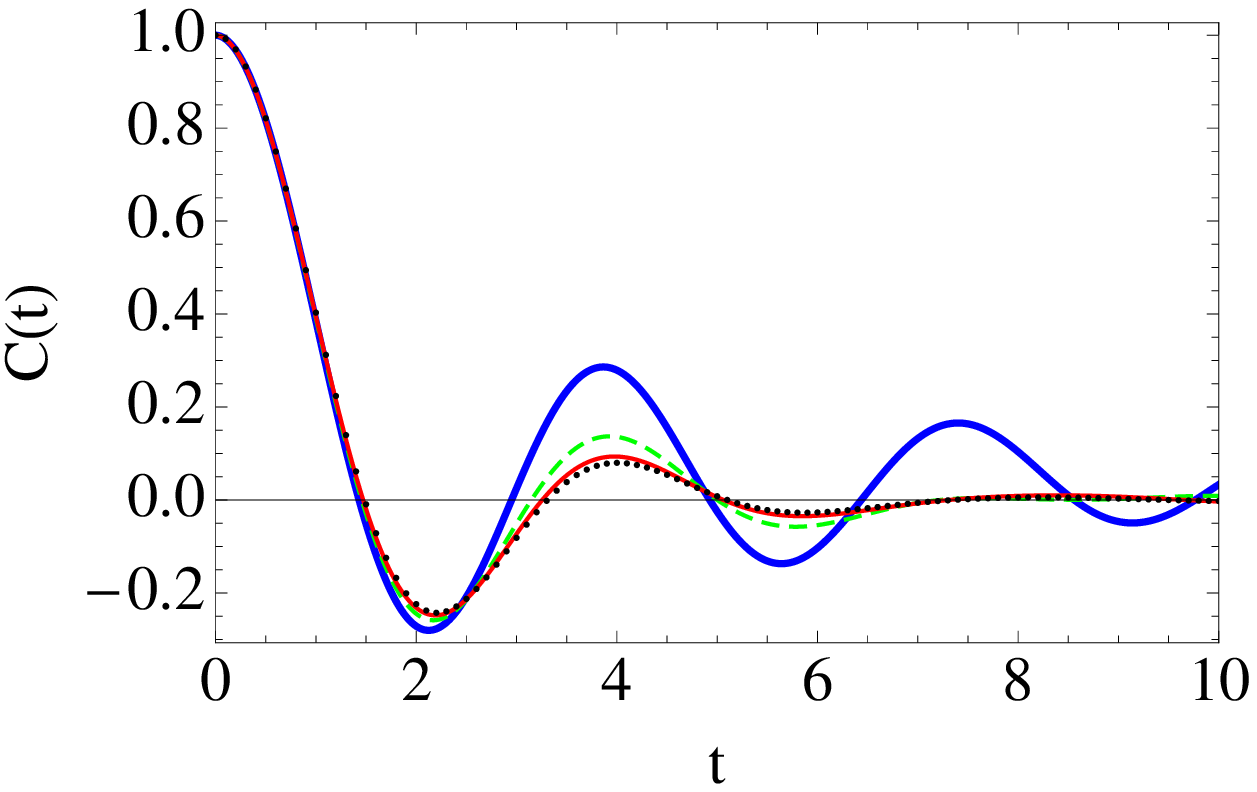,width=8cm } }
\put(0, 0){ \epsfig{file=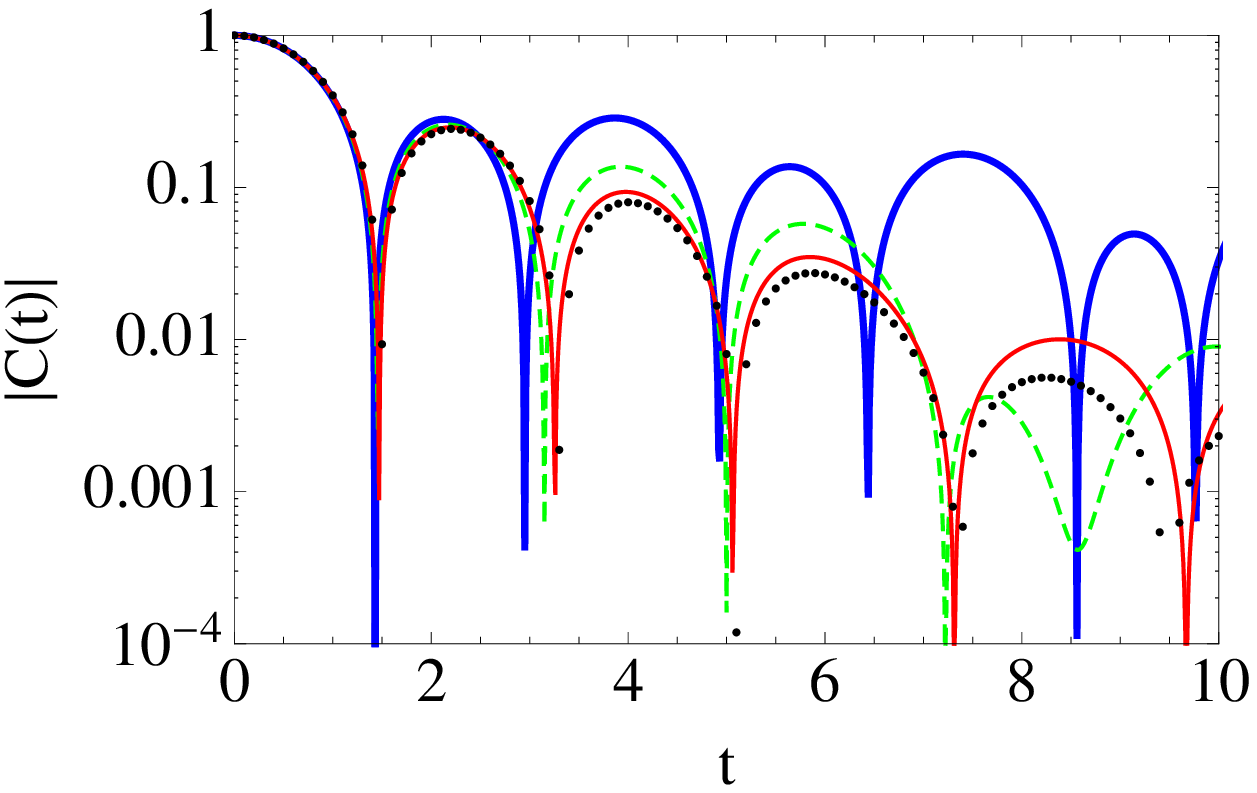,width=8cm } }

\put(0,102) { {\bf(a) }}
\put(0,48) { {\bf(b) }}
}
\end{picture}
\caption{   (Color online) Correlation functions $C(t)$ for spin chains with nearest-neighbor interactions.  Black dots - classical spins, blue thick solid lines - spins 1/2,  green dashed lines -  spins 1, and red thin solid lines -spins 5/2. The first three chains consist of 12 spins. The last one consists of 9 spins. Coupling constant for classical spins are $J^x_{mn} = J^y_{mn} =-0.41$ and $J^z_{mn}=0.82$. Coupling constants for quantum spins are rescaled as described in the text. (a) Linear-scale plot. (b) Semi-logarithmic plot. } 
\label{fig-s}
\end{figure}

In Fig. \ref{fig-s}, we present correlation functions $C(t)$ for a classical spin chain and for  quantum spin chains with $S=1/2,\ 1$ and $5/2$.  All chains consist of 12 spins except the spin-5/2 chain which consists of 9 spins.  For the classical Hamiltonian, we take the nearest-neighbor coupling constants  $J^z_{mn}=0.82$ and $J^x_{mn} = J^y_{mn} =-0.41$, while for the quantum Hamiltonians we divide the above values by $\sqrt{S(S+1)}$. As explained in Section~\ref{general}, this is done to match the characteristic timescales and the initial behavior for quantum and classical lattices.

We first notice in Fig. \ref{fig-s} a significant difference between the classical correlation function and the correlation function for the spin-1/2 chain. The latter function is rather unusual, because it does not exhibit clear long-time behavior of form (\ref{long-t}). Instead, its long-time behaviour appears to be a modification of Eq.(\ref{long-t}) that relaxes not to zero but rather to a non-zero ``baseline'' that itself is slowly approaching zero, possibly, exponentially.  We have checked that this behavior is not a finite-size effect by obtaining the same behavior in the same time range for a chain of 24 spins 1/2.  This appears to be  a transitional case anticipated in Ref.~\cite{Fine-04}, when two long-time relaxation modes, oscillatory and monotonic, decay with nearly the same exponential rate $\gamma$ and hence coexist. The above behavior may also be a peculiar manifestation of the integrability of spin-1/2 chains with the nearest-neighbor interaction. However, other examples of integrable spin-1/2 chains considered in Ref.~\cite{Fine-04} exhibited the generic long-time behavior of form (\ref{long-t}).

At the same time, we observe in Fig. \ref{fig-s} that the correlation function for the spin-1 chain is already quite close to the correlation function for classical spins, while, for spin-5/2 chain, the agreement with the classical result is excellent. 

\begin{figure*} \setlength{\unitlength}{0.1cm}
\begin{picture}(176 , 106 )
{
\put(0, 54){ \epsfig{file=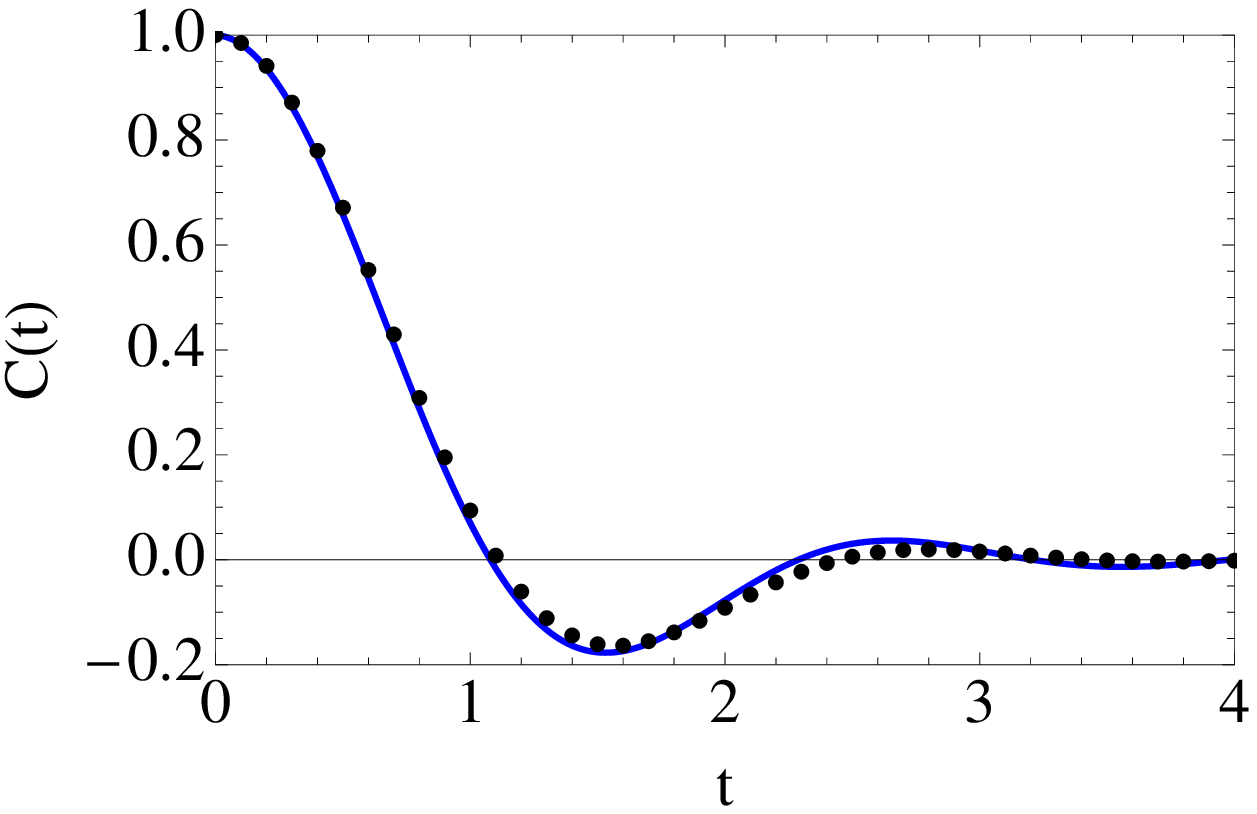,width=8cm } }
\put(0, 0){ \epsfig{file=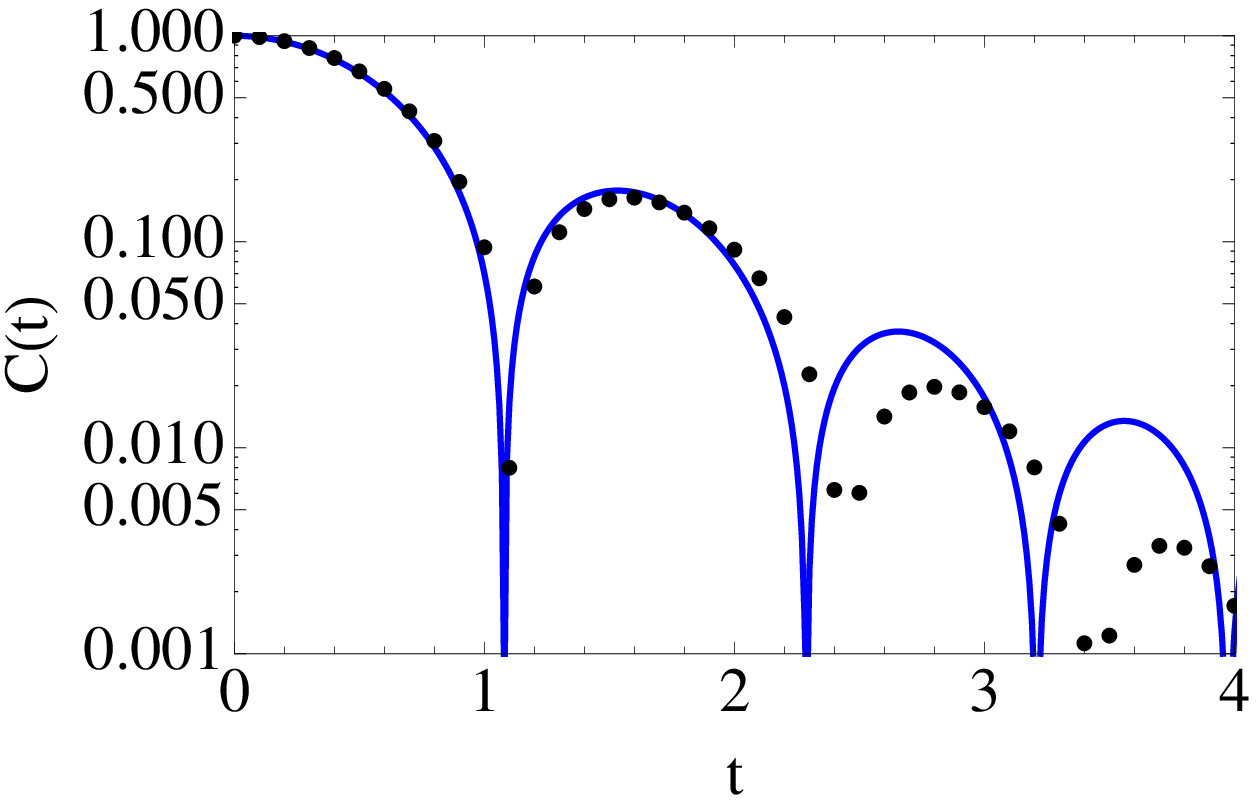,width=8cm } }

\put(-2,102) { {\bf(a) }}
\put(-2,48) { {\bf(b) }}

\put(88, 54){ \epsfig{file=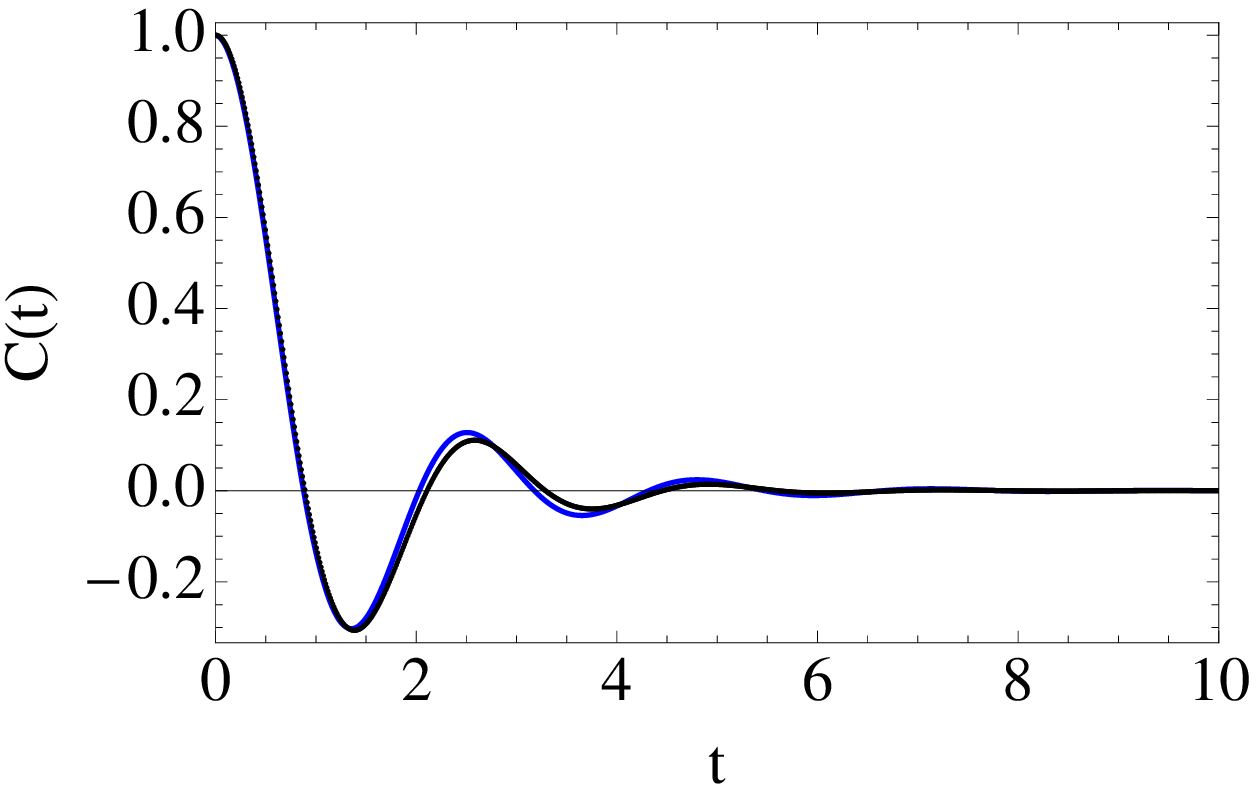,width=8cm } }
\put(88, 0){ \epsfig{file=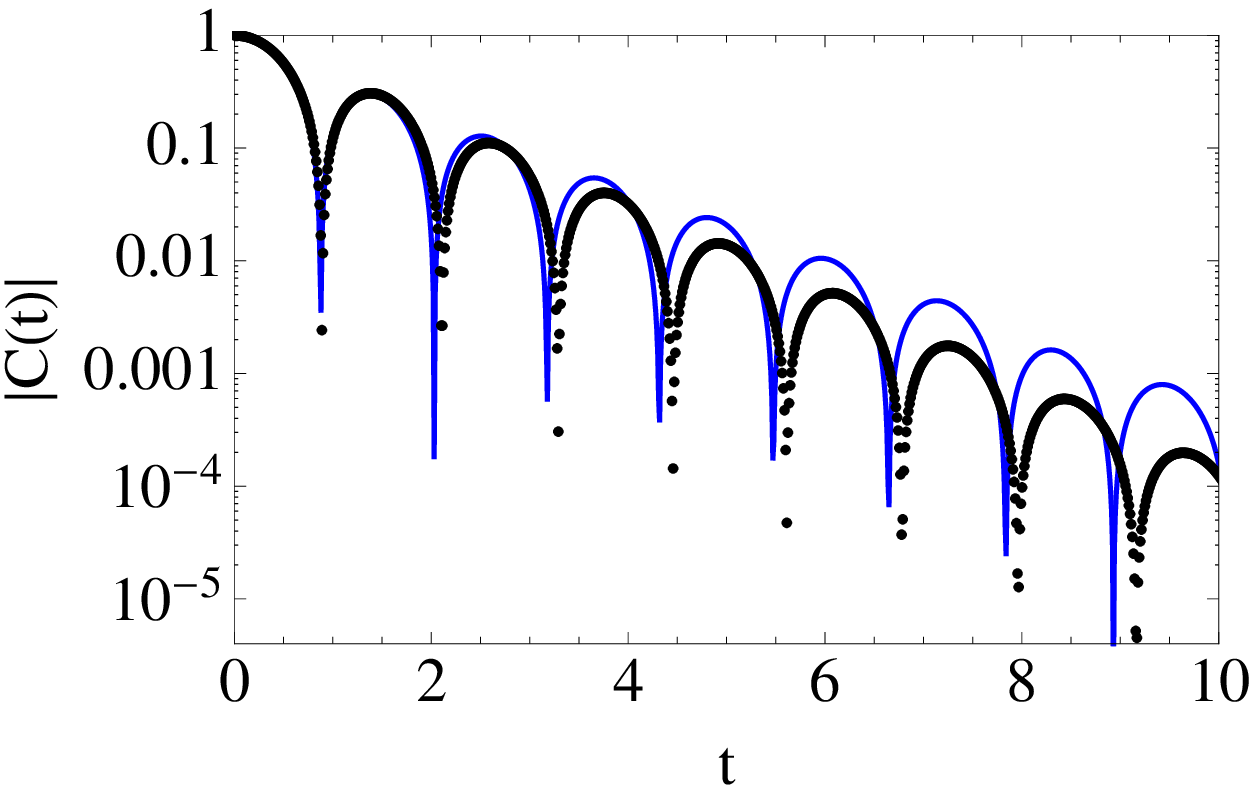,width=8cm } }

\put(86,102) { {\bf(c) }}
\put(86,48) { {\bf(d) }}

}
\end{picture}
\caption{  (Color online) Correlation functions $C(t)$ for $5 \times 5$ square spin lattices with nearest-neighbor interactions.  Black dots - classical spins, blue solid lines - spins 1/2. Coupling constants for classical spins are: (a,b) $J^x_{mn} = J^y_{mn} =-0.41$ and $J^z_{mn}=0.82$; (c,d) $J^x_{mn} = 0$, $J^y_{mn} =-1$ and $J^z_{mn}=1$. Coupling constants for quantum spins are rescaled as described in the text. (a,c) Linear-scale plots. (b,d) Semi-logarithmic plots.
} 
\label{fig-2D}
\end{figure*}

In Fig.~\ref{fig-2D}, we compare FIDs for two-dimensional $5 \times 5$ square lattices of spins 1/2 and classical spins. We considered two classical spin Hamiltonians with the nearest-neighbor coupling constants either $J^x_{mn}= J^y_{mn} = -0.41$, $ J^z_{mn} =0.82$,  or $J^x_{mn}=0$,  $J^y_{mn} = -1$, $ J^z_{mn} = 1$. For the corresponding quantum lattices the constants are rescaled as before. We notice that the agreement between the correlation functions for the quantum and classical lattices is as good as for $^{19}$F FID in CaF$_2$ with [100] direction of magnetic field.

\section{Concluding discussion}
\label{discussion}

As mentioned in the introduction, the relevance of classical spin simulations for describing quantum spin dynamics at high temperatures has been appreciated for long time. However, this relevance was generally believed to be ``semi-quantitative"\cite{Tang-92}. In the present article, we have shown that, for certain class of quantum spin lattices, classical simulations give quantitatively accurate result. The empirical requirements for the overall quantitative agreement on linear-scale plots (i.e. excluding exponentially vanishing long-time behavior dominating on the semi-logarithmic plots) appear to be the following:

(i) The system should be translationally invariant, and the correlation functions of interest should decay on the fastest natural time scale of the system $\tau$ given by Eq.(\ref{tau}). (For some correlation functions decaying on the time scale slower than $\tau$ --- for example, FID in the presence of exchange narrowing \cite{Abragam-61} --- the quantitative agreement may also be good, but the present article contains no numerical  investigations of such cases.)

(ii) For spin-1/2 lattices, each spin should have at least 4 strongly interacting neighbors. In the case of magnetic dipolar or similar kinds of interaction involving varying coupling constants, the effective number of strongly interacting neighbors $n_{eff}$ can be defined using the participation ratio of the neighbors in mean-squared fluctuations of the local field:
\begin{equation}
\label{neff}
n_{eff} = \frac{\left( \sum_n \left\langle h_{mn}^2 \right\rangle \right)^2}{\sum_n \left\langle h_{mn}^2 \right\rangle^2} = 
\frac{
\left[  
\sum_n \left( {J^x_{mn}}^{\!\!\!\! 2}  +   {J^y_{mn}}^{\!\!\!\! 2}   +   {J^z_{mn}}^{\!\!\!\! 2}  \right) 
 \right]^2
 }
 {  
\sum_n  \left( {J^x_{mn}}^{\!\!\!\! 2}  +   {J^y_{mn}}^{\!\!\!\! 2}   +   {J^z_{mn}}^{\!\!\!\! 2}    \right)^2
 }.
\end{equation}
Here $\left\langle h_{mn}^2 \right\rangle$ is the contributions of the $n$th spin to the mean-squared local field fluctuations experienced by the $m$th spin.  For $^{19}$F FIDs in CaF$_2$ the values of $n_{eff}$ are 4.9, 9.1 and 22.2 for [100], [110] and [111] magnetic field directions respectively. For the lattices with nearest-neighbor interactions, the above formula just gives the number of the nearest neighbors. 
 Thus, the threshold $n_{eff}=4$ is based on the satisfactory quantitative agreement for the square lattices and for $^{19}$F FID in CaF$_2$ with magnetic field along the [100] direction.

(iii) For lattices of spins $S$ with $ S \geq 1$, two strong neighbors should be sufficient. (This judgment is made on the basis of the results presented in Fig.~\ref{fig-s}. It is also consistent with findings of Ref.~\cite{Gade-66}.)

As far as the long-time behavior of correlation functions is concerned, then classical simulations can also be used to accurately predict the constants of this behavior for spin-1/2 systems with $n_{eff} \gtrsim 9$. (See the calculations of $^{19}$F FIDs in CaF$_2$ for [110] and [111] directions of magnetic field). However, in examples with $n_{eff} \leq 5$ noticeable discrepancies remain, which increase with decreasing $n_{eff}$, and become particularly dramatic for the FID in the spin-1/2 chain presented in Fig.~\ref{fig-s}. 

In a broader context, the correspondence between classical and quantum spin dynamics touches on the important phenomenon of chaos.  Classical chaos is defined as exponential sensitivity to small perturbations of phase space trajectories. Classical spin lattices are generically chaotic\cite{deWijn-12,deWijn-13,Elsayed-14}. In contrast, quantum spin systems do not have phase space trajectories. It has recently been shown by us that nonintegrable lattices of spins 1/2 exhibit power-law rather than exponential  sensitivity to small perturbations \cite{Fine-14}. The difference can be observed in the behavior of NMR magic echoes (also known as Loschmidt echoes). In this case the exponential sensitivity emerges with the increase of the values of quantum spins\cite{Elsayed-14A}. It, therefore, appears that the unifying aspect of classical and quantum chaos that leads to the quantitative agreement between classical and quantum relaxation described in the present article is the ergodicity of the underlying dynamics rather than the exponential sensitivity of the system to small perturbations. Ergodicity is compatible with both exponential and power-law sensitivities to small perturbations.

In this article we did not consider disordered lattices and hence avoided dealing with the issues of glassy dynamics and many-body quantum localization, both of which can suppress ergodicity.  Comparison of classical and quantum relaxation in the presence of disorder remains an interesting issue, which requires further investigation. 

Finally, we speculate that microscopic classical simulations are likely to be quantitatively accurate not only for doing first-principles calculations of NMR free induction decays in solids but also for a broader class of quantum problems, where individual quantum microscopic degrees of freedom are far from the classical limit.

The authors are grateful to bwGRiD project \cite{bwgrid} for computational resources.

\bibliographystyle{apsrev4-1}
%

\end{document}